\documentclass{emulateapj}
\usepackage{graphicx, amsmath}
\usepackage{natbib}
\usepackage{float}

\usepackage{silence}
\defcitealias{Stroke1967}{G. W. Stroke, "Diffraction Gratings" in Handbuch der Physik (1967)}

\shorttitle{Grism Design and NIRCam Grisms}
\shortauthors{Deen et al.}

\begin{document}
\title{A Grism Design Review and the As-Built Performance of the Silicon Grisms for JWST-NIRCam}
\author{Casey P. Deen}
\affiliation{Max Planck Institut f\"ur Extraterrestrische Physik, Giessenbachstrasse 1, Garching bei M\"unchen, Deutschland}
\affiliation{Max Planck Institut f\"ur Astronomie, K\"onigstuhl 17, D-69117
Heidelberg, Deutschland}
\affiliation{Department of Astronomy University of Texas at Austin, 1 University Station, 78712, Austin, TX, USA}
\author{Michael Gully-Santiago}
\affiliation{Department of Astronomy University of Texas at Austin, 1 University Station, 78712, Austin, TX, USA}
\affiliation{Kavli Institute for Astronomy and Astrophysics, Peking University, 5 Yiheyuan Road, Haidian District, Beijing 100871, P. R. China}
\author{Weisong Wang}
\affiliation{University of Dayton, Department of Electrical and Computer Engineering, Kettering Laboratories 341, 300 College Park, Dayton, 45469, Ohio, USA}
\author{Jasmina Pozderac}
\affiliation{Applied Physics Laboratory, Johns Hopkins University, 11100 Johns Hopkins Road
Laurel, MD 20723-6099, USA}
\author{Douglas J. Mar}
\affiliation{Liquidia Technologies, 419 Davis Dr, Morrisville, 27560, North Carolina, USA}
\author{Daniel T. Jaffe}
\affiliation{Department of Astronomy University of Texas at Austin, 1 University Station, 78712, Austin, TX, USA}

\begin{abstract}
Grisms are dispersive transmission optics that find their most frequent use in instruments that combine imaging and spectroscopy.  This application is particularly popular in the infrared where imagers frequently have a cold pupil in their optical path that is a suitable location for a dispersive element.  In particular, several recent and planned space experiments make use of grisms in slit-less
spectrographs capable of multi-object spectroscopy. We present an astronomer-oriented general purpose introduction to grisms and their use in current and future astronomical instruments.  We present a simple, step-by-step procedure for adding a grism spectroscopy capability to an existing imager design.  This procedure serves as an introduction to a discussion of the device performance requirements for grisms, focusing in particular on the problems of lithographically patterned silicon devices, the most effective grism technology for the 1.1-8 micron range.  We begin by summarizing the manufacturing process of monolithic silicon gratings.  We follow this with a report in detail on the as-built performance of parts constructed for a significant new space application, the NIRCam instrument on JWST and compare these measurements to the requirements. 
\end{abstract}

\maketitle

\section{Introduction}
\label{sec:intro}

Grisms are transmission gratings combined with prisms to produce a single diffractive
optical element.  Manufacturing techniques for grisms include optically contacting 
planar transmission gratings to a refractive prism, ruling or machining grooves 
directly into a prism-shaped substrate, or patterning onto and etching the grooves 
into the substrate.  Grisms are extremely versatile and useful devices
for moderate resolution spectroscopy ($R=\frac{\lambda}{\Delta\lambda}\sim
100<R<10,000$), especially at infrared wavelengths.  In the infrared, grisms
compete with highly efficient dispersive prisms at the low spectral resolution
($R\sim100$) end of their resolving power range.  At higher resolving powers,
surface-relief reflection gratings have historically formed the alternative to
grisms.  Recently, however, instrument designs for $R=1,000-22,000$ at
$1-2.5\mu$m have begun to make use of volume phase holographic gratings (VPH) in
transmission as well \citep{Tamura2006, Insausti2008, Wilson2010, Yuk2010}.

In astronomical spectrometers, the main advantage of grisms over transmissive gratings 
is their ability to disperse the light while
sending the blaze wavelength forward along its original, undeviated path.
Either conventional surface-relief transmission gratings or volume phase holographic gratings can be used in a grism, but VPH gratings typically use double-prisms to straighten the beam \citep{Ebizuka2011}. 
Optical designers value
straight-ahead transmissive dispersers because they offer the possibility 
of mechanically simple, compact systems.  The main
appeal of the undeviated path, however, stems from the opportunity it presents
to make dual use of infrared imaging systems as
spectrometers.  Most infrared imagers include optics to reimage the telescope
pupil onto a cold stop to minimize the incidence on the detector of thermal
radiation from warm surfaces in the telescope and foreoptics.  In an IR reimager, if the beam is adequately collimated at the location of the pupil stop, then an optical designer may press a filter wheel near the pupil stop into service as a grating
carrier as well.  If the first 
focal plane lies within the cold part of the system, a wheel at this location
containing slits and field stops make the system fully convertible from imaging
to spectroscopy with minimal effort.  Examples of such instruments include NIRC 
on the Keck telescope \citep{Matthews1994}, CONICA at the VLT \citep{Lenzen1998}, 
NSFCAM at the IRTF \citep{Rayner1998}, TIMMI$-$2 for the ESO 3.6m telescope
\citep{Reimann2000}, MIMIR for the Lowell Observatory Perkins telescope 
\citep{Clemens2007}, FLITECAM \citep{Smith2006} and FORCAST \citep{Keller2000_SPIE} for SOFIA, and WFC3 on 
Hubble \citep{Baggett2007, Kuntschner2010}.

In this paper we begin by giving a design guide for grism spectrometers.  Much
of this material can be derived from first principles or exists elsewhere in
some basic form, but there is a need for a clearly laid out procedure for preliminary 
grism spectrometer design to permit instrument builders to find the right
path along which to begin the optimization of more detailed computer-aided
designs.  The designs discussed in this section of the paper naturally lead to
requirements for the grism characteristics.  The choice of grism type and the
design of grism-based spectrometers are constrained by the material properites
of the prism, the grating, and the adhesive (if bonded), and by the limitations
of various groove production techniques.  Most astronomical grism spectrometers
currently in use rely upon dispersive elements produced by directly ruling
grooves into refractive materials or by imprinting grooves into a compliant medium
and then hardening the material on the surface of a prism.  Recently,
lithographic techniques have progressed significantly and to the point where it
is possible to pattern and etch grooves into crystalline silicon subtrates with
enough accuracy for infrared grism applications \citep{GullySantiago2010,
Vitali2008, Mar2006, Kaeufl1998}.  In our previous detailed paper on Si grisms
\citep{Mar2009}, we presented results on our first-generation devices destined
for use in the FORCAST mid-IR camera on SOFIA \citep{Adams2010}.  The Mar paper
focuses on design and manufacturing issues for silicon grisms at the device
level.  In this paper, we present a new generation of Si grisms produced for use
in the $2-5\mu$m range in the NIRCam instrument on JWST \citep{Jaffe2008,
Greene2010}. These devices have reached a new level of accuracy and optical
performance.  By showing detailed metrology of the etched silicon grisms, we
illustrate both the manufacturing tolerances and how different types of defects
affect various aspects of the disperser performance.

\section{Grism Spectrometer Design}
\label{sec:design}

\subsection{Strawman Design of a Long-slit Grism System}
\begin{figure}
 \begin{center}
    \includegraphics[width=8cm]{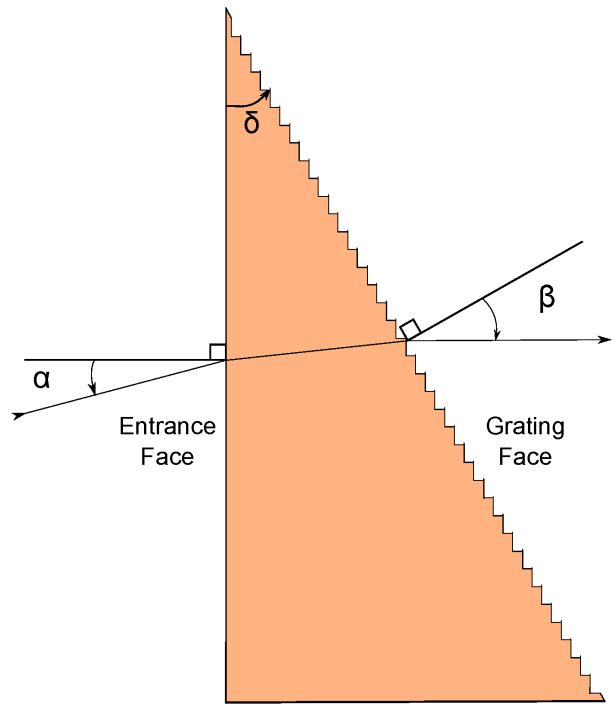}
  \end{center}
  \caption{Grism schematic.  In this example, light enters the dielectric material through the
  flat face on the left and is dispersed as it exits through the grating surface.}
  \label{fig:grism_schematic}
\end{figure}

We can derive the grating equation and dispersion relation for grisms easily from first principles.  It is useful, however, to present a set of equations and an illustration that make clear the relevant angles and the correct sense in which to measure them.  If we consider a refracting prism with index $n$ and opening angle $\delta$ (also known as the ``apex angle'') and having an optically flat entrance surface and a grating with groove spacing $\sigma$ on the exit surface, we can derive the generalized form of the grating equation for grisms by determining the angle $\beta$ from the grating normal on the exit face, where the phase shift between adjacent grooves, given an incident angle relative to the surface normal of the entrance face $\alpha$, is an integral number $m$ of waves.  Figure \ref{fig:grism_schematic} illustrates the geometry.  (Note that $\alpha$ is measured counter-clockwise and $\beta$ clockwise from their respective surface normals.)

\begin{equation} \label{eq:grism_eqn}
\frac{m\lambda}{\sigma} = n\left(\lambda\right) \sin\left(\delta -
\sin^{-1}\left(\frac{\sin\left(\alpha\right)}{n\left(\lambda\right)}\right)\right) - \sin \beta
\end{equation}

For most grisms at most infrared wavelengths, the change in refractive index with wavelength is small compared to the dispersion of the grating itself, so $n\left(\lambda\right)=n$ can be held constant until accuracies better than a few percent are required.  One important consequence of equation \ref{eq:grism_eqn} is that under certain conditions the direction of the output beam with respect to that of the input beam
at a given wavelength is only very weakly dependent on the angle of incidence of the grism:
\begin{equation}
\label{eq:input_output}
\alpha - \beta \simeq \text{constant}
\end{equation}
Take, for example, a typical set of parameters for a first-order silicon ($n=3.41$) grism by assuming an opening angle $10^\circ$ and a groove spacing of $4.779$ $\mu$m.  Place this grism in the beam with the flat face toward the incident light.  A beam at normal incidence on this grism ($\alpha=0^\circ$) will have $\beta=10.000^\circ$ if $\lambda=2.0\mu$m, and will therefore emerge from the grism undeviated from its original path.  If we now tilt this grism so that $\alpha=1.000^\circ$, the grating equation tells us that $\beta=9.001^\circ$, that is, the beam deviates now by $\frac{1}{1000}$ of a degree from its initial direction.  For a camera with a focal length of $100$mm, this angular displacement translates into a displacement of $1.75\mu$m, or about $\frac{1}{10}$ of a pixel for a typical infrared focal plane array.  This insensitivity to the orientation of the grism with respect to the incoming beam means that even a moderately well-engineered grism/filter wheel will produce a very repeatable spectral 
format on the array.

\begin{figure}
 \begin{center}
    \includegraphics[width=8cm]{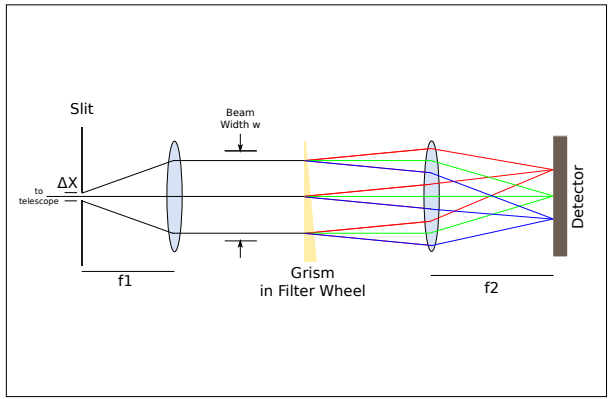}
  \end{center}
  \caption{Generic Grism Instrument.  $\Delta x_{\rm slit}$ is the physical width of the slit (or, in
  the case of a slitless instrument, the image size). The grism sits at or near the pupil
  formed by the collimator (focal length $f_1$).  The camera (focal length $f_2$) images the 
  dispersed light onto the detector.  For use as an imager, the slit and grism are removed
  from the beam.}
  \label{fig:grism_instrument}
\end{figure}

Since the most common use of grisms is to provide a spectroscopic capability to an imaging system, we use this type of instrument as a worked example of grism spectrometer design.  The straw man imaging system has a cold focal plane, a collimator that produces a pupil where a stop can be inserted, and 
then a camera that re-images the first focal plane onto the detector.  A generic grism instrument (see Figure \ref{fig:grism_instrument}) has an entrance slit of width $\Delta x_{\rm slit}$ placed at the initial cold focal plane of the imaging
system.  (Note that, in the case of a slitless spectrograph, the quantity $\Delta x_{\rm slit}$ represents the physical size of the images in the focal plane).  We assume for simplicity that the grism goes at an image of the telescope pupil. The camera then re-images the now-dispersed focal plane onto the detector.

In our design example, we assume that the imaging camera already exists or that
its design is complete.  This assumption then fixes several critical parameters:
the width of the pupil $W$, the focal length of the collimator, $f_1$, the focal
length of the camera, $f_2$, the pixel size in the focal plane array $\Delta
X_{\rm pix}$, and the number of unvignetted pixels across the field, $N$.  If
the camera has a good achromatic design and the desired range of $\beta$ lies
within the range of field angles for the imager, the optical system should
perform well as a spectrometer.  We are going to calculate here a preliminary
design.  Detailed calculations would need to account for the geometric effects
of prism dispersion and for aberrations, as well as for diffraction at the slit.
 For simplicity, we ignore the dispersion of the prism which is usually small compared to the dispersion of the transmission grating part of the grism.  The diffraction limited
spectral resolution is equal to the number of waves of delay between the top and
bottom edges of the beam as they pass through the grism:

\begin{equation} \label{eq:Rdiff}
R_{\rm diff} = \frac{\lambda}{\Delta \lambda} = (n-1) \frac{W}{\lambda}
\tan{\delta}
\end{equation}

In a straight-through optical system like this one, the grating will not cause
any anamorphic distortion and a monochromatic point source at the first focal
plane will appear round in the image plane so the physical size of a
diffraction limited spectral resolution element will be equal to that of a
diffraction limited spatial resolution element.  The four elements we have
control over are the slit width $\Delta x_{\rm slit}$, opening angle $\delta$,
the groove spacing $\sigma$, and the blaze angle $\xi$.

It turns out that adjusting the parameters in the correct order greatly
simplifies the exercise of arriving at an optimal preliminary design.  The
width of the slit determines the sampling in the spectral direction.  The
number of pixels across the slit is just

\begin{equation} \label{eq:npixslit}
n_{\rm slit} = \left(\frac{f_1}{f_2}\right) \left(\frac{\Delta x_{\rm
slit}}{\Delta x_{\rm pix}}\right)
\end{equation}

Setting $\Delta x_{\rm slit}$, in this context, determines not only the
sampling but also the angular width of the slit on the sky since the imaging
camera design has already fixed the plate scale.  In an approximation valid as
long as the size of the wavelength interval on the detector array $\Delta
\lambda_{\rm tot}$ is modest compared to the wavelength, the opening angle of
the grism determines $\Delta \lambda_{\rm tot}$:

\begin{equation} \label{eq:dlambda}
\Delta \lambda_{\rm tot} = \lambda \frac{N}{f_2} \left[ \frac{\Delta x_{\rm pix}}{\left(n-1\right) \tan \delta} \right]
\end{equation}

The combination of the slit width and the opening angle of the prism determine
$R$, the resolving power of the system.  In this preliminary design, we
approximate by assuming that the geometric slit width determines the resolving
power.  A more detailed design would need to take into account diffraction
effects, since infrared systems often operate with slit widths only a few times
larger than the diffraction limit, and the effects of convolution with the
response of a small number of finite-sized pixels.  Note that geometric ray
tracing programs also ignore these effects.

\begin{equation} \label{eq:resolvingpower}
 R = R_{\rm diff} \frac{f_1 \lambda}{W \Delta x_{\rm slit}} = \left(n-1\right)
\tan \delta \frac{f_1}{\Delta x_{\rm slit}} 
\end{equation}

The resolving power is equal to the diffraction-limited resolving power given
in equation \ref{eq:Rdiff} divided by the width of the slit in units of
diffraction-limited spot sizes.  Note here how the resolving power depends
directly on $\left(n-1\right)$ and inversely on the slit size.  The presence of
the factor $\left(n-1\right)$ in equation \ref{eq:resolvingpower} 
means that the choice of a high index material can permit
substantially higher resolving powers at a given slit width or substantially
larger slits on the sky at a given resolving power.  The factor of
$\left(n-1\right)$ is $\sim 6$ times larger for Si than for low index materials
like $\rm CaF_2$.

The choice of groove spacing now determines the wavelength along 
the undeviated path.  For
light that passes undeviated through the grism, we have $\beta =
\delta - \alpha$.  For a given order and orientation of the grism with respect
to the incident beam, $\alpha$, we can then use equation \ref{eq:grism_eqn} to
derive the appropriate value of $\sigma$ to make a particular wavelength
$\lambda_0$ pass through the grism undeviated.  Since we have shown above that
the value of $\alpha - \beta$ is very insensitive to the grating orientation,
we can get a good estimate of the correct value for $\sigma$, for a given order m, by assuming
$\alpha = 0$.  In this case,

\begin{equation} \label{eq:sigma}
 \sigma = \frac{m \lambda_0}{\left(n-1\right) \sin \delta}
\end{equation}

\subsection{Choice of Order}
\label{sec:order}
The usually correct choice in designing a grism instrument is to use the grism
in first order.  The resolving power slit-width product is independent of grism
order and the angular dispersion is independent of the chosen order once $\tan
\delta$ and $\lambda_0$ are fixed.  For the near-IR atmospheric windows, first
order has the advantages that one can choose to use wavelengths only at the very
peak of the blaze function and that blocking light in higher orders is
relatively easy.  There are, however, a few reasons to choose to use grisms in
orders $m > 1$.  The most common of these is when the science requirements call
for the ability to take spectra in several different infrared windows where one
might for example, choose a grating that operated in first order in the
$K$--band and second order in the $J$--band.  In higher orders, the
polarization effects of the grating decrease so that one can trade free
spectral range for a more polarization$-$independent performance.  Finally,
for the Si gratings we discuss in \S \ref{sec:actual_grisms}, there are
throughput advantages in higher order.  In the silicon grating manufacturing
process, the coarser grooves present at higher order will have a smaller
fraction of their lengths blocked by the etch stops necessary in production.  In
addition, it will be easier to deposit multi-layer anti-reflection coatings on
physically larger grooves.

\subsection{Blaze}
\label{sec:blaze}
\begin{figure}
 \begin{center}
    \includegraphics[width=8cm]{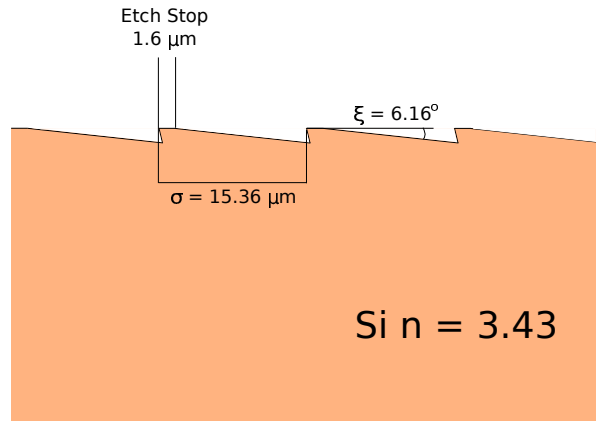}
  \end{center}
  \caption{Groove cross-section schematic for the JWST A6-I grism for the NIRCam 
  instrument.}
  \label{fig:groove_xsection}
\end{figure}

Equations \ref{eq:grism_eqn} - \ref{eq:sigma} set all the critical parameters
that determine valid solutions for the direction taken by the dispersed light
emerging from the grism.  The power distribution between the different orders
and within a given order as a function of wavelength, however, depends on the
groove geometry: the shape of the individual grooves, their orientation with
respect to the grating normal, and the width of the grooves relative to the
size of the spacing between them.  Derivation of the blaze function is complex
and requires detailed calculation of the electromagnetic behavior of the groove
structure, taking into account the geometry, the material properties, and the
polarization of the radiation \citep{Neviere1990, Neviere1991}.  The most
common sorts of infrared grisms for astronomy either employ flat-faceted
grooves that fill the groove intervals and have $90^{\circ}$ corners or, in the
case of micro-machined Si grisms, flat-faceted grooves with $70^{\circ}$
vertices and small (filling-factor $\sim 10\%$) lands parallel to the
grating surface in between the facets (Figure \ref{fig:groove_xsection}).
Detailed calculations of the blaze functions in first order for these two
geometries are available in the literature \citep{Neviere1991, Mar2009}.  These
calculations show that, even in first order, the peak of the actual blaze
function lies close to the analytically calculated blaze wavelength for
gratings with opening angles $<45^{\circ}$.  Therefore, a grating used at
normal incidence on the flat entrance should be blazed so that the facets on
the grooved side are parallel to the entrance face $\left(\xi=\delta\right)$ if
the undeviated wavelength is to be at the peak of the blaze.

\subsection{Cross-Dispersed Grism Designs}

Higher order ``echellette'' grisms allow for high spectral resolution but reasonable numbers of spectral points across a single order.  In a camera/spectrometer system, one can cover larger swaths of wavelength at higher spectral resolution, albeit at the cost of slit length, by inserting a low-order grism cross-disperser into the beam immediate preceding or following the echellette \citep{Dekker1988}.  The ability to produce high quality coarse grooves in silicon using micro-lithographic techniques now makes such cross-dispersed systems possible in the infrared \citep{Ennico2006}.

Cross-dispersed grism spectrometer design depends on a more complex set of trades than are necessary for designing a long-slit spectrometer with a single grism.  The result of our analytic design procedure is therefore somewhat less accurate than in the single grism case but still provides a good way to make trades and a good way to start a cross-dispersed design.  Once again, we start by assuming the existence of an imager with a pre-determined collimated beam size and collimator and camera focal length.  The insensitivity of the grisms to tilt also makes it difficult to tune the central wavelength, so we will assume that all of our analytic designs seek to provide continuous wavelength coverage over a given spectral interval.  The simplest way to look at the design is to see that the total number of unvignetted pixels in the focal plane fixes the product of a series of quantities:
\begin{equation} \label{eq:scale_factor}
\eta N^2 = \frac{\Delta \lambda_{\rm tot}}{\lambda_C} \times R \times n_{\rm slit} \times \frac{\zeta_{\rm min}}{\theta_{\rm pix}}
\end{equation}
where $\eta$ is a scale factor to allow for the variation in order length and order separation, as well as in the dispersion of the cross-disperser across the wavelength range and to allow for blank pixels between orders.  A value of $\eta$ between 0.25 and 0.5 is usually reasonable.  $\Delta \lambda_{\rm tot}$ is the full (fixed) wavelength range.  $\lambda_C$ is the central wavelength.  $R_{\rm slit}$ is the slit-limited resolving power from (\ref{eq:resolvingpower}), $\zeta_{\rm min}$ is the minimum desired slit length, and 
$\theta_{\rm pix}$ is the projected size of a pixel, both in units of angle on the sky. With so many variables to juggle, it is hard to know how to proceed with the initial design.  One sensible path is to first fix the sampling $n_{\rm slit}$ based either on a desired limit on pixel smoothing of the slit point spread function or on the desire for an particular angular size of the slit width.  One then chooses the wavelength range, since this is often quantized for you by the boundaries of the atmospheric windows.  After that, it is a matter of trading slit length against resolving power.

When following the above recipe, there are two additional constraints that are worth keeping in mind: (1) You cannot get an arbitrarily high value of resolving power for a given value of the beam diameter, $W$.  The maximum value of the prism opening angle $\delta$ is physically limited by the mechanical depth of the clearance along the line of sight at the position of the filter wheel.  For lithographically produced Si gratings with their $70^{\circ}$ groove vertices, you are further limited by shadowing of the groove profiles (see \citet{Mar2009} for a quantitative assessment of this effect). (2) You cannot choose an arbitrarily large wavelength range without sacrificing throughput since the blaze function of the cross-disperser in first order will limit the total range and since expanded range can only be achieved by reshaping the grooves at the expense of peak efficiency.

Having fixed most of the quantities in equation \ref{eq:scale_factor}, we see that the tradeoff between $\zeta_{\rm min}$ and $R_{\rm slit}$ is really a trade between $\tan \delta$ and $\zeta_{\rm min}$.  We can set a minimum value for the slit separation by requiring the slit to be some number of point-source image lengths long.  This requirement will place an upper limit on the value of $\tan \delta$.  Now, to meet the requirement for the continuous wavelength coverage, we need to choose the order in a way that keeps the entire free spectral range on the detector.  This will be true for a central order $m_C$ such that:

\begin{equation} \label{eq:central_order}
 m_c \simeq \eta^{\frac{1}{2}} \frac{f_2}{N} \times \left(n-1\right) \times \frac{\tan \delta}{\Delta x_{\rm pix}}
\end{equation}

We can now use equation \ref{eq:sigma} to obtain the grism groove spacing for the echellette grism.  The approximate number of orders in the echellogram is given by the total wavelength coverage divided by the free spectral range in the central order, $\left(\frac{\Delta \lambda_{\rm tot} m_C}{\lambda_C}\right)$.

\section{Silicon Grisms}
\subsection{Process}
The choice of grism material must take into account the transmission properties and refractive index of the substrate prism material and the manufacturability of the blazed grating surface.  Since most applications in infrared astronomy involve cryogenic systems, the ability of the grism to survive repeated thermal cycles is also important.  Astronomers have fielded instruments with glass/epoxy, KRS5, ZnSe, Si, and Ge grisms, among others.  Apart from the advantage in slit width--resolving power product that Si enjoys as a result of its large refractive index, silicon has a number of significant advantages as a grism material in the near-IR: excellent transmission at $1.2$ -- $8  \mu$m \citep{Briggs1950} and good transmission at wavelengths beyond $20 \mu$m \citep{Lord1952}, hardness, machinability, vacuum and cryogenic compatibility, radiation hardness, thermal shock immunity, insolubility, the existence of good materials for anti-reflection coating (at least at $\lambda < 8 \mu$m), and a well-understood fabrication process for monolithic devices.  Over the past 15 years, the University of Texas grating group has developed and refined techniques for producing high-quality grisms made from crystalline silicon.  Parts of two previous refereed papers describe the production process \citep{Marsh2007, Mar2009}.  We update that discussion here based on experience over the past few years and expand it to include explanatory material for an astronomical, rather than a silicon processing audience.

\begin{figure}
 \begin{center}
    \includegraphics[width=8cm]{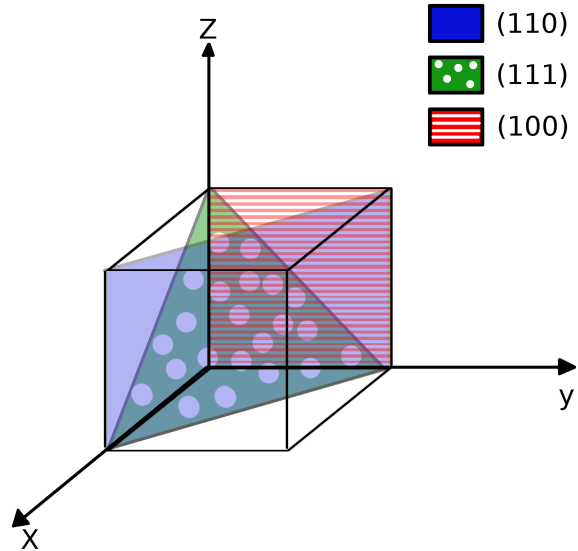}
  \end{center}
  \caption{A cartoon representation of the various crystal planes referenced in the processing description.}
   \label{fig:Crystallography_Schematic}
\end{figure}

The basic process for producing blazed gratings in crystalline silicon relies on the peculiarities of silicon etch chemistry for its success \citep{Tsang1975}.  By using anisotropic etchants, chemicals that remove silicon more quickly along one crystal axis than another, it is possible to produce pyramids, v-grooves, and other shapes at the surface of silicon crystals.  The $\left\lbrace111\right\rbrace$ family of crystal planes intersect at a $70^{\circ}$ angle.  For those unfamiliar with Miller index crystallography notation, Figure \ref{fig:Crystallography_Schematic} shows a cartoon representation of the various crystal planes discussed in our procedure.  An aqueous solution of potassium hydroxide (KOH), prepared at the correct strength and temperature, can remove material perpendicular to the $\left(100\right)$ plane by up to 100 times faster than it removes material perpendicular to the $\left(111\right)$ plane \citep{Seidel1990}.  By properly patterning a prepared surface, one can take advantage of the crystalline properties of Si and the chemical behavior of KOH to produce triangular grooves with $70^{\circ}$ vertices.

We begin with a $100$ mm diameter boule of high resistivity monocrystalline silicon.  The low impurity content of this material results in a smaller number of local disruptions of the Si crystal lattice and fewer scattering or absorption sites in the bulk material than one would encounter in lower resistivity Si.  Commercial boules are pulled perpendicular to either the $\left(100\right)$ or $\left(111\right)$ planes and oriented to an accuracy of order one degree.  In the production of grisms, we usually begin with $\left(111\right)$ material because this orientation optimizes the processing and shaping of the finished optic from the center of the boule, minimizing waste material.

The azimuthal orientation determines the direction of the grooves across a crystal face since these grooves can only be produced along the intersection of a $\left(111\right)$ and a $\left(100\right)$ plane.  Trial and error experimentation shows us that we can produce clean grooves with no discontinuities if we know the azimuthal orientation of the boule around the $\left[111\right]$ vector to about $\pm 0.05^{\circ}$.  We precisely determine the initial orientation of the silicon using x-ray crystallography and then mark the azimuthal angle by machining a ``clocking'' flat into the sides of the cylindrical boule.  The crystal structure itself (the intersection angle between two $\left(111\right)$ planes) determines the $70^{\circ}$ opening angle of the groove vertices.  What remains is to orient the exposed crystal face with respect to the $\left(100\right)$ plane to determine the angle between that face and the bisectors of the groove vertices.  This orientation determines the blaze of the transmission grating, which would be $54.7^{\circ}$ if the grating face and $\left(100\right)$ plane were coincident.  It also determines the relative lengths of the two sides of the groove vertex (with the two sides having equal length at $54.7^{\circ}$).  We then saw a series of $8-20$ mm thick disks from the boule, canted at the appropriate angle to produce the desired blaze.  Each disk has a portion of the clocking flat on one side to show the future direction of the grooves.

We polish each disk from the original boule optically flat using the chemical mechanical planarization process employed to produce wafers for the microelectronics industry.  This process uses grit diluted in a slurry of chemical etchant.  As the grit mechanically removes material from the surface, the etchant removes further material and cleans out the disruptions in the crystal lattice caused by the grit.  Specular surfaces with $\frac{\lambda}{10}$ surface accuracy at visible wavelengths are easily achievable with this method.

In order to produce the desired grooves, we must direct the anisotropic etching process.  We do this by producing a set of stripes in a KOH-resistant material that can serve as a barrier to etching across the $\left(100\right)$ plane.  We begin by vapor-deposition of a thin ($\sim 500$ nm) layer of the passivation material Si$_3$N$_4$.  Silicon nitride is extremely resistant to KOH and effectively stops etching on all Si surfaces it covers.  We spin-coat this passivation layer with an additional thin layer of binary photoresist.  The manufacturers formulate this UV--sensitive emulsion to be non-linear so that it produces well-determined transitions between exposed and unexposed material.

The desired pattern on the planar surface is a series of thin stripes.  VLSI technology companies can now produce large (20 cm scale) photomasks with absolute accuracies of $\sim 10$ nm.  The commercially obtained photomask consists of a series of parallel chrome stripes on quartz.  We orient the stripes to the correct azimuthal angle by aligning to the clocking
flat on the Si disk and then contact the mask to the disk surface.  After exposing this assembly to UV light, we develop (and thereby remove) the exposed resist, leaving behind the desired stripes in the resist, evenly spaced at the groove constant of the grating with regions of Si$_3$N$_4$ in between.  Using a reactive ion etcher, we remove the exposed silicon nitride. Subsequent removal of the protective stripes of photoresist using an organic solvent leaves us with a flat Si disk with even spaced stripes of silicon nitride running parallel to the intersection of the $\left(100\right)$ and $\left(111\right)$ planes.

With the stripes in place, immersion in KOH produces the desired triangular grooves.  Etching proceeds $30-100$ times faster across the $\left(100\right)$ plane than across the  $\left(111\right)$ plane.  When material is removed to the point that a $\left(111\right)$ plane lies exposed along the edge of a Si$_3$N$_4$ strip, the rate slows dramatically, allowing the material between stripes to etch down to a single vertex.  The finite $\left\langle100\right\rangle$/$ \left\langle 111\right\rangle$ anisotropic etch ratio has two notable effects: a small amount of undercutting of the silicon nitride strips reduces their width and the longer etch time at the top of the groove compared to the bottom effectively rotates the blaze angle by a small amount.  For an anisotropic etch ratio of 60, this rotation is about $0.3^{\circ}$ \citep{Mar2009}.

\begin{figure}
 \begin{center}
    \includegraphics[width=8cm]{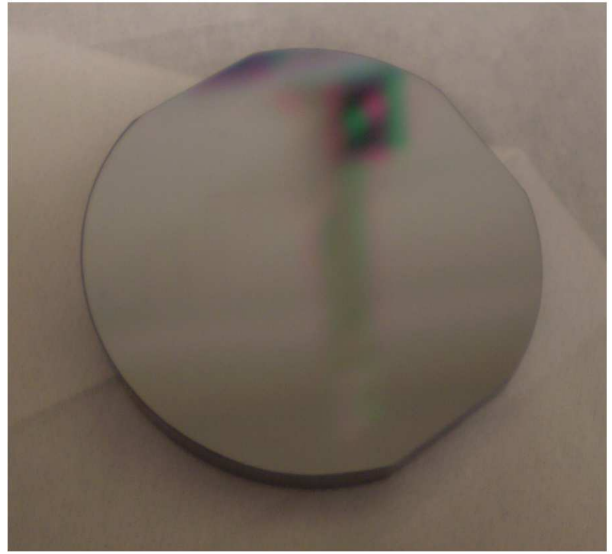}
  \end{center}
  \caption{Photograph of the completed but not yet coated JWST grism UT-A6-I.  The part is $\sim45$ mm in diameter.  The grooved surface faces the viewer.  The clocking flat is on the upper right corner}
   \label{fig:part}
\end{figure}

Once the etching process produces the grooves (Figure \ref{fig:groove_xsection}), we need to cut and polish the back side of the silicon disk at the appropriate angle to produce a prism with the desired value of $\delta$ (Figure(\ref{fig:part})).  We can also apply an anti-reflection coat both the flat face and the grooved face of the grism to optimize the transmission over the desired operating range.

\section{Requirements, Metrology, Results}
\label{sec:actual_grisms}
This section discusses the specific manufacturing requirements for grisms.  We first discuss generic requirements relevant to all grism materials and all production methods and then discuss 
some that are only relevant to lithographic Si grisms.  Careful metrology is a part both of the process development and of the verification that devices will be able to perform at the required level.  We therefore integrate into the discussion of requirements a discussion of measurement techniques and actual results.  As a test case, we report the measurement results for a grating produced as one of a batch of six for use in the NIRCam instrument on JWST \citep{Jaffe2008, GullySantiago2010}.

In the standard camera/grism configuration, a polychromatic collimated beam, either from the slit or from the entire field in a slitless configuration, passes through a grism inserted at or near the pupil (Figure \ref{fig:grism_instrument}).  For each constituent wavelength in the input beam, an ideal grism will transmit an input planar wavefront and divide it into a series of planar wavefronts with output 
directions determined by the wavelength.  The multiple solutions to the grating equation ( \ref{eq:grism_eqn}) determine the directions of these wavefronts.  The interaction of the wave with the shape of the individual grooves determines the distribution of power among the directions allowed by the grating equation.  Along with the aberrations and losses introduced elsewhere in the optical chain, irregularities in the grism distort the input plane wave.  These irregularities divide by their source:  distortions in the shapes of the prism surfaces or deviations of the groove positions, by their scale: from the size of the whole device to scales smaller than a wavelength, and by their character: random or repetitive, scale invariant or single-sized.

\begin{figure}[H]
 \begin{center}
    \includegraphics[width=8cm]{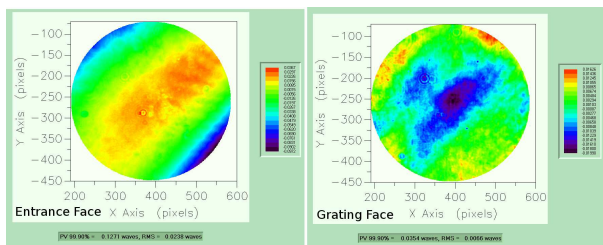}
  \end{center}
  \caption{Interferometric measurements of the JWST NIRCam grism UT-A6-I.  Measurements give the phase error (in units of waves at 632 nm) measured in reflection at 633 nm over a 42mm beam.  (left) Interferogram of the flat entrance face. (right) Interferogram of the grating grooves measured at the Littrow angle.}
   \label{fig:grism_flatness}
\end{figure}

The entrance face of the prism is the first phase-altering element encountered by the incident wavefronts.  A physical deviation of the entrance face in the propagation direction, $\epsilon_{ef}$, leads to a phase error of 2$\pi (n(\lambda)-1) \epsilon_{ef}/\lambda$ where n($\lambda$) is the refractive index of the prism. For a reflective surface, the standard requirement is for a peak-to-valley flatness of $\lambda$/4 for diffraction-limited performance.  For transmission through silicon, a deviation of $\epsilon_{ef}$  induces a phase change of 4.8$\pi \times \epsilon_{ef}$ /$\lambda$, comparable to the 4$\pi \times \epsilon_{ef}$/$\lambda$ one would see in reflection.  This larger change leads to a flatness requirement of $\epsilon_{ef}<\lambda/5$ for a single silicon surface. (For lower index materials, the (n-1) term means that the flatness requirement becomes rapidly less stringent.) The left-hand panel of figure \ref{fig:grism_flatness} shows an interferogram of the front surface of the NIRCam grism UT-A6-I (all measurements are of this particular part, unless otherwise specified).  The reflected phase error for this part is 0.127$\lambda$ peak to valley at 633nm over a 42 mm diameter.  This corresponds to a transmitted wavefront error of  $\sim \lambda$/20 at 2 $\mu$m.  Over the 31 mm pupil, the deviation is about half this size.  

Flatness errors in the parent surface of the grooved side of the grism, usually introduced in the
polishing step, also enter into the overall wavefront budget for the grooved face of the grism since the grooves are ruled or etched into or bonded onto this surface.  Since polishing errors are not random (rounding down at the edges of the piece, for example), we need to recognize that the flatness part of the error in the grooved face can add linearly with the error in the flat entrance face, rather than in quadrature.  Since the opening angles of grisms are typically modest, the flatness of the parent surface for the grating enters into the error budget with a factor of cos($\delta$) $\simeq$ 1, the two surfaces of the grism should therefore be polished to a peak-to-valley flatness of $\lambda$/10 at the operating
wavelength.

The other source of distortion at the grooved surface arises from the placement of the grooves along that surface.  Geometry, however, makes the placement of the grooves along this parent surface much more forgiving than the flatness of the surface itself.  Physical deviations of the groove positions along the grating surface, $\epsilon_{gs}$, lead to phase errors of

\begin{equation} \label{eq:phase_error}
 \frac{2\pi \left(n-1 \right)} {\lambda} sin\left(\delta \right) \times \epsilon_{gs}
\end{equation}

For a silicon grism like UT-A6-I (sin$\delta$= 6.16 degrees), deviations in the placement of the grooves results in errors in the direction of propagation that are a factor of 9 smaller than the errors caused by flatness deviations of comparable size.  Despite this favorable circumstance, it is the groove placement and finish that can cause the most problems for grisms.  We can isolate the phase properties of the groove face from those of the entrance, albeit combining errors in the flatness of this face and in the proper placement of the grooves, by measuring the surface error in reflection.   As with the entrance face, we have a flatness requirement of $\epsilon_{gs}<\lambda/10$ at the observing wavelength in vacuo for Si and correspondingly less stringent by the ratio of (n($\lambda$)-1) for lower index materials.  The right-hand panel of Figure \ref{fig:grism_flatness} shows the interferometer results for UT-A6-I.  The peak to valley deviation of the grooves is 0.035 waves at 633nm on all scales, about a factor of  ~30 better than required at the operating wavelength of 3.5 $\mu$m.

The imperfections in groove placement on different scales give rise to defects in the wave front (and therefore in the spectral point spread function), many of which have traditional labels derived from the long history of ruled grating manufacture.  We therefore divide our discussion of the groove errors by scale and character.  Errors on scales approaching the full size of the pupil produce the classic aberrations: sphere, coma, astigmatism etc.  This large-scale power (at low amplitude) goes into distorting and broadening the core of the spectral point spread function and is therefore most important when the slit size is close to the diffraction limit.  Errors on intermediate and small scales spread power away from the core of the monochromatic image and reduce the efficiency of the grating.  For small errors with a random distribution in spatial frequency, the degradation in the actual intensity $\eta$, relative to the maximum possible intensity for a diffraction-limited spectral image, taking into account all other loss factors, $\eta_o$ goes as 

\begin{equation} \label{eq:random_error}
\frac{\eta}{\eta_o} = e^{- \left( \frac{2\pi \left(n(\lambda)-1 \right)} {\lambda}  \sin\left(\delta \right) \left(\epsilon_{gs}\right) \right)^2}
\end{equation}

This is an equation for reflective losses in \citet{Marechal1970} modified to account for phase errors in transmission \citep{Keller2000} ,where ${ \epsilon_{gs}}$ sin($\delta$) is the standard deviation of the groove position in physical units along the direction of propagation. In a blazed grating the smallest spatial frequency of the error corresponds to the groove size.  Random displacement therefore spreads the lost power throughout the blaze pattern of the grating.  The loss from random errors reaches 10\% ($\eta/\eta_o$)= 0.9) when $\epsilon_{gs}$=$\lambda$/[19.4*(n($\lambda$)-1)sin($\delta$)], or $\epsilon_{gs}$=$\lambda$/5.0 for Si grisms like UT-A6-I with $\delta$=6.16$^\circ$.  The measurements in the right panel of Figure \ref{fig:grism_flatness} show that the random errors and large-scale errors combined are at a level many times below this at the operating wavelength of this device.

\begin{figure}[H]
 \begin{center}
    \includegraphics[width=8cm]{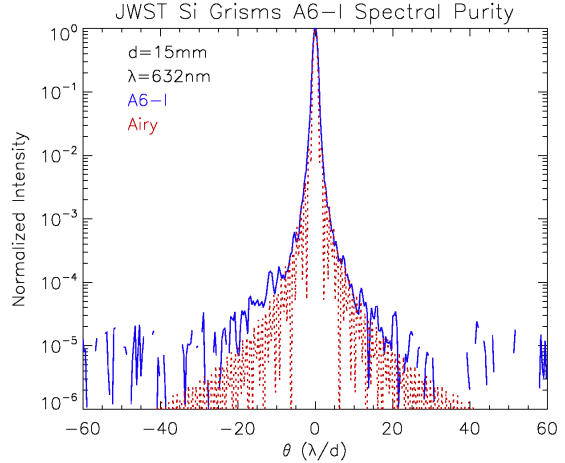}
  \end{center}
  \caption{Logarithmic plot of the monochromatic spectral point spread function of the grating surface for grism UT-A6-I.  This measurement was made in reflection at optical wavelengths.  The blue line shows a one-d image of a 633 nm laser source in reflection from the grating.  The red dashed line is the theoretical diffraction-limited psf}
   \label{fig:grism_psf}
\end{figure}

Repetitive errors are the most pernicious form of moderate-scale errors since they concentrate the power caused by the phase variation into discrete features (ghosts).   Sinusoidal variations in groove position with amplitude $\epsilon_{rep}$ (measured along the grating surface) produce first-order Rowland ghosts with intensities, \textit{I}, relative to that of the parent line, \textit{I}$_o$, of
\begin{equation} \label{eq:rowland_ghost}
\frac{I}{I_o} =  \left[ \frac{2\pi \left(n-1 \right) \epsilon_{rep} sin\delta} {\lambda} \right]^2 
\end{equation}
(from \citetalias{Stroke1967}, as modified to account for the difference between a surface-relief grating used in reflection and a grism made with a material with refractive index $n$) The  need to measure accurate line fluxes or equivalent widths for weak lines drives the requirements on ghosts.  The sky subtraction process we use to remove telluric emission lines will also effectively eliminate the ghosts of these lines. Since astronomical emission lines are sparsely distributed, we can easily recognize their ghosts.   Ghosts of astronomical absorption lines impose the most stringent constraints.  If we wish to be able to measure equivalent widths to an accuracy of 0.2\% of the continuum, the ghost levels must be smaller than this value.  The deepest astronomical lines at a resolving power of a few thousand are about 50\% of the continuum.  The maximum ghost relative intensity should therefore be less than 0.4\%.  As an example, for a silicon grism with $\delta=6.16^\circ$, this intensity imposes a limit on the amplitude of any repetitive error of $\epsilon_{gs}<\lambda/25.7$.  This ratio corresponds to an amplitude of $<$45  nm if we are working at the cut-on wavelength of 1.15 $\mu$m.  Figure \ref{fig:grism_psf} shows the spectral point spread function of the grating surface for grism UT-A6-I.  There are no apparent ghosts down to the 10$^{-4}$ level.

Micro-roughness of the groove surfaces also causes loss.  In machined gratings, this roughness arises from tool marks or imperfections in substrate itself.  In lithographically produced gratings, this roughness arises both as a result of localized defects in the crystal and from process issues such as H$_2$ bubbles that form on the groove surface as the etching is under way.  The scattered light from these defects spreads over large solid angles.  The fractional loss due to these defects due to an rms roughness $\epsilon_{rms}$ is:

\begin{equation} \label{eq:roughness}
\frac{I}{I_o} =  \left[ \frac{2\pi \left(n(\lambda)-1 \right) \epsilon_{rms}} {\lambda} \right]^2 
\end{equation}

We can determine local roughness in groove surfaces using atomic force microscopy \citep{Mar2009} or non-contacting profilometry \citep{Wang2010}.  For lithographically produced silicon grooves with properties similar to those in UT-A6-I, the rms roughnesses reported in these papers were 1.7-2.5 nm.  The roughness losses were therefore 0.1\% or less all the way down to the 1.15$\mu$m Si cutoff.

In lithographically produced gratings, large-scale defects in the form of missing groove sections or pits or hillocks in the grating surface can result from problems during the lithographic pattern transfer, from dust flecks on the substrate surface, from pinhole defects or scratches in the passivation layer, or from large crystal dislocations.  These large-scale defects will scatter flux over angular scales related to the inverse of their linear dimensions.  For defects a few microns in size, this corresponds to 5-10$^\circ$ scales.  These defects should occupy no more than 10\% of the surface if they are to remain a small contributor to the intensity losses. Inspection of UT-A6-I and similar parts shows that the area covered by such defects is usually $<$1\%.

\subsection{Groove Blockage in Lithographic Si Gratings}

One source of loss peculiar to lithographically produced Si grisms arises from the crystal geometry and the manufacturing process.  In transmission, the $70^{\circ}$ vertex angle of the Si groove profiles means that the unused groove surface is in the path of a portion of the incident beam.  This partial blockage results in a double loss, equal to a $1-\left(1-b\right)^2 \simeq 2b$.  The extra loss is a manifestation of Babinet`s principle, where the blockage itself accounts for one factor of $b$, and the narrower width of the blazed grooves permits additional power to leak into adjacent orders accounting for the other factor of $b$.  The fact that some of the light incident on the blocked part of the grooves may continue in the forward direction complicates matters and needs to be analyzed with full-wave electromagnetic analysis but most likely only modifies the destination of the lost radiation and not the size of the loss.  The fraction of the beam that does not hit a blazed groove surface due 
to the $70^{\circ}$ vertex angle is easy to calculate for the most commonly used grism geometry
where the light enters the device normal to the flat entrance face ($\alpha = 0$) and where the 
grating grooves are blazed for the undeviated beam.   This purely analytic approach is not strictly correct for low-order gratings but the results are consistent with a more sophisticated electromagnetic model \citep{Mar2009}. The fractional blockage $b$ is.

\begin{equation} \label{eq:blockage}
 b = \cot 70.5^{\circ} \times \tan \delta = 0.35 \tan \delta
\end{equation}

Equation \ref{eq:blockage} implies that the blockage problem should become quite small for grisms
with small opening angles.  Even at small values of $\delta$, however,  
some portion of the incoming beam will not strike the blazed groove surface because the etching process requires us to leave small, flat intervals in the grating plane that serve as etch stops (Figure \ref{fig:groove_xsection}).  With our current manufacturing technique, the loss from the etch-stop flats is $\sim 10\%$ but we can reduce that value using precision electron beam lithographic patterning \citep{GullySantiago2014}.  The minimum strip width for contact lithography is $~2\mu$m while concerns about undercutting and breakthrough of the etch barriers place a minimum size of the flats 
produced by electron-beam lithography of a few hundred nanometers, or $3-5\%$ of the groove spacing, whichever is larger.  At larger $\delta$ angles, the groove ``lip'', the the protrusion into the beam resulting from the $70^{\circ}$ vertex angle, responsible for the loss and reduction of the size of the flats has no effect.
Figure 6 of \citet{Mar2009} plots the loss as a function of blaze angle for one reasonable geometry.  The loss reaches $20\%$ for $\delta > 16^{\circ}$.  Silicon grism designs therefore must limit the prism opening angle to less than this.

\subsection{Blaze Variations Due to Anisotropic Etching Variations}

Another possible source of groove position error in etched silicon gratings may come into play if there are variations in the groove angle across the grating.  In lithographically patterned and etched gratings, the edge of the etch-stop determines the groove position.  Starting from this position, a deviation in groove angle (blaze) moves the phase center of the groove off of the regular spacing needed for perfect performance.  The effect of this kind of error is not distinguishable from other sources of phase deviation. For a grating with the geometry of UT-A6-I and a required peak-to-valley deviation for random errors of $<\lambda/10$ at $\lambda= 1.15 \mu$m, the maximum allowable deviation in blaze angle is 0.8$^\circ$.  The most likely cause of such a change would be a variation in the anisotropic etch ratio across the piece.  Changes in the rate at which KOH etches across the $<$100$>$ plane versus how fast it etches across the $<$111$>$ plane changes
the orientation of the groove blaze with respect to the grating surface \citep{Mar2009}.  The ratio would have to change from a typical value of 60:1 to below 20:1.  Measured variations in the ratio on a given part, however, are considerably less than this.  The interferogram of the grooved surface of UT-A6-I, with its total peak-to-valley error of 0.035 waves confirms that this effect can be small for lithographic gratings.  

\subsection{Bulk and Surface Absorptive and Reflective Loss}

Bulk absorption is a potential problem
for grisms, as for any other transmissive optical elements.  
Typical path lengths through grisms, however, are very comparable to those through
many lenses, so designers can apply the same considerations they apply to lens 
materials to the suitability of grism substrates.  As a rule, one seeks to
avoid materials with more than a few percent absorption over the maximum
path-length through the grism.  The major difference with the case of lenses
is the way in which the material will be processed.  For grisms,  
manufacturability through ruling, imprinting, or lithography will drive 
material choice, along with refractive index and
bulk absorption.  The high-resistivity float-zone silicon used in the JWST grisms
has very modest loss (Mar et al. 2009 and references therein) and bulk loss is therefore
a negligible factor at 1.15 $\mu$m $< \lambda < 8 \mu$m for prisms with dimensions
comparable to those of our JWST devices.

A second-order consequence of attenuation in the bulk grism material is, in effect, to apodize
the beam in the pupil.  The side of the pupil with a longer path length has a weaker field than
the short path length side and this variation alters the shape of the spectral point
spread function.  Mar et al. (2009) examined this effect and found that it
was negligible until the attenuation in the grism became unacceptably large.

\begin{figure}[H]
 \begin{center}
\includegraphics[width=8cm]{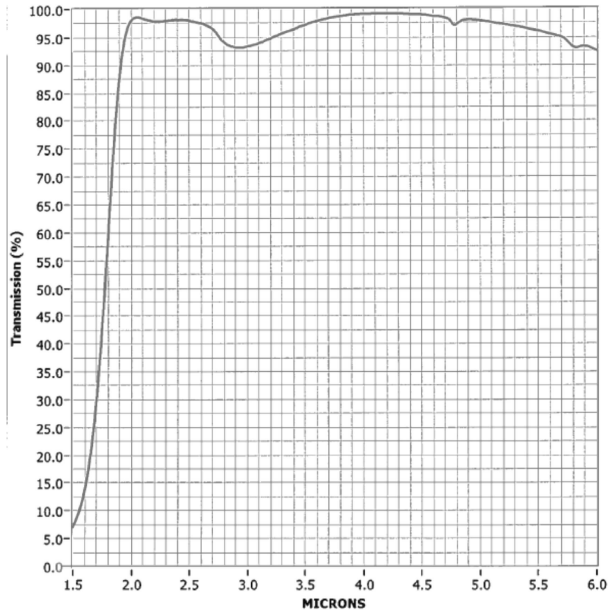}
  \end{center}
  \caption{Transmission of a silicon witness sample.  The sample has been antireflection 
coated on both sides with the same coating as JWST grim A6-I (D. Kelly, Steward Observatory, Univ. of Arizona, personal communication)}
  \label{fig:twosided_witness}
\end{figure}

In addition to the bulk absorption, there is a Fresnel loss at both the entrance surface and
the exit surface of the grism.  Because of the dispersion at the exit face, there is no 
constructive interference as would be the case for a double-sided disk and the loss is 
simply the concatenated loss at the two surfaces.  For Si, the large refractive index that
is such a boon for prisms in other respects becomes a liability.  Over the near-IR, the
typical loss is 30\% per surface.  Passage through two uncoated surfaces leads then to an
absolute maximum efficiency of 50\%.  
Fortunately, there are good broad-band antireflection coatings for Si that can reduce the loss
at a single surface to $<2$\% over the entire 2 to 6 $\mu$m range \citep{GullySantiago2010}.  Figure (\ref{fig:twosided_witness}) shows a double-side coated silicon witness sample
produced along with the coatings on JWST grism A6-I.  Typical transmission is better than 97\% from
2 to 5 $\mu$m.  It is somewhat unclear, however, the extent to which the performance of the witness
sample is relevant to the reduction of Fresnel losses at the grooved surface.  The sharp shape
acute angles, and small scale sizes of etched Si grism grooves all conspire to make evaporative 
deposition of precise and uniform thicknesses somewhat problematic. The topography of the coating may have change the mechanical stresses on the part during thermal cycling.  Layer-by-layer surface 
profilometry of coated grooves would be able to provide data that could tell us how effective
the groove coatings are, but such tests have not been carried out by the coating manufacturers.  

\subsection{End-to-End Efficiency of Silicon Grisms in the mid-IR}

\begin{figure}[H]
 \begin{center}
    \includegraphics[width=8cm]{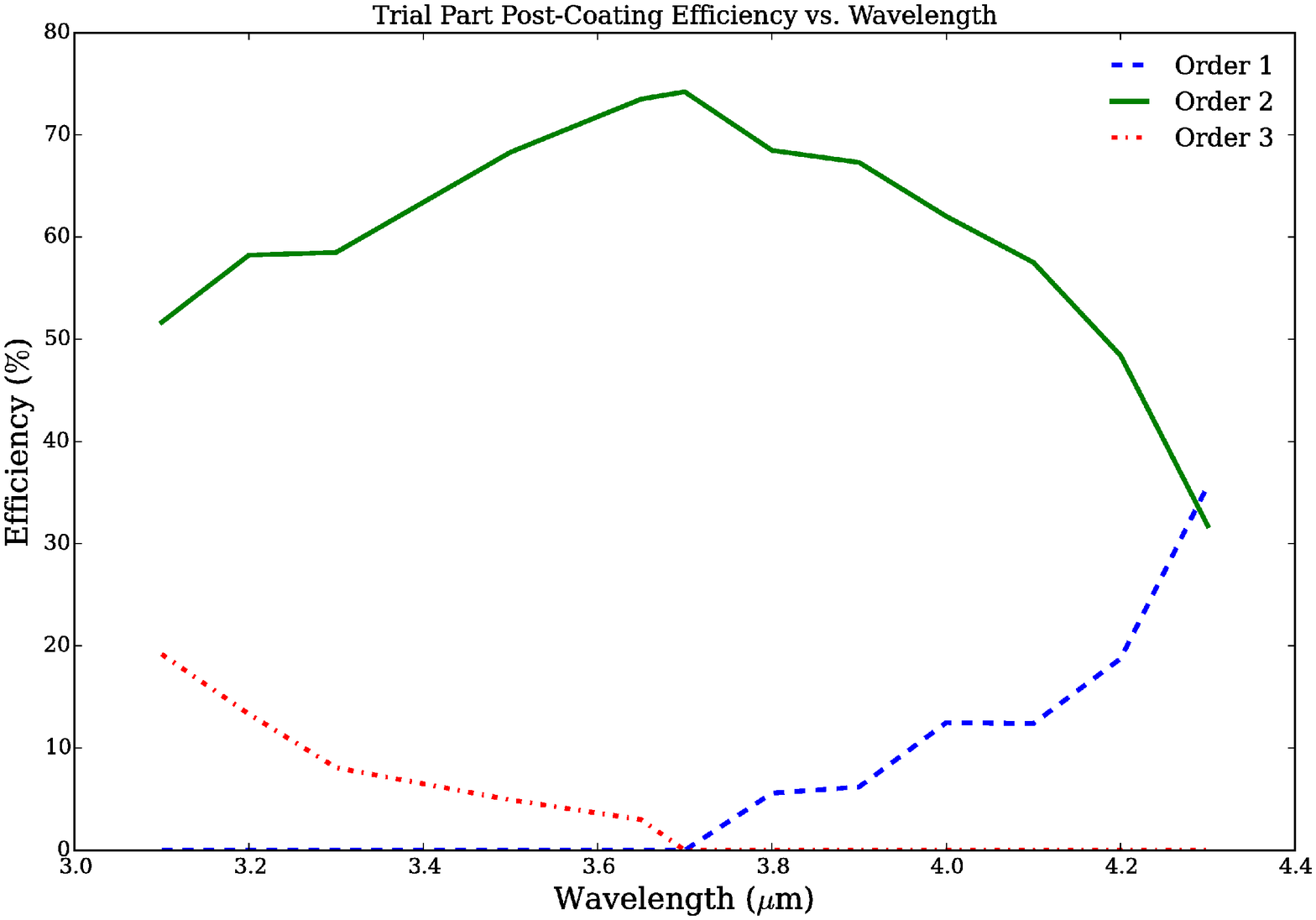}
  \end{center}
  \caption{End-to-end efficiency of the JWST "trial part".  This part was physically identical
to the flight gratings except that the grooves are blazed at a larger angle.}
   \label{fig:trial_part}
\end{figure}

Contractual and scheduling issues in the production of the grisms for JWST meant that we were unable
to measure a component-level optical efficiency for the A6I or the other completed 
devices.  We substitute here a measure of the "trial part", a device moved through
our process ahead of the flight parts to provide a final test of each production test.
This part was physically identical to the flight parts in all respects except the 
initial angle of the grating surface with respect to the (111) crystal plane.  
Where we cut the flight part disks
to have the grating surface at an angle of 6.16$^o$ with respect to 
the (111) plane, the trial part disk had
an angle of 11$^o$ between the grating surface and the (111) plane.  This difference is
important because, when we then cut the substrate to form a prism, instead of the facets on 
the exit face being parallel to the flat entrance face, they were tilted by about 5$^o$ in
the opposite direction from the opening angle $\delta$ (Fig. \ref{fig:grism_schematic}). 
The net result of this tilt was to place the blaze direction at lower values of $\beta$ and
so to increase the blaze wavelength in a given order (Equation \ref{eq:grism_eqn}). 
We patterned this part using the JWST mask to produce the same groove
constant as on the flight parts.  With the larger blaze angle, however, 
the blaze peak in second order lies close to the
wavelength of the flight parts when these are in first order.  Figure \ref{fig:trial_part} shows the efficiency of this part as a function of wavelength.  The peak efficiency in second order is $\sim$75\%.
The peak efficiency of the flight
parts should be comparable, although their first-order blaze functions should be significantly
broader.

\begin{figure}[H]
 \begin{center}
    \includegraphics[width=8cm]{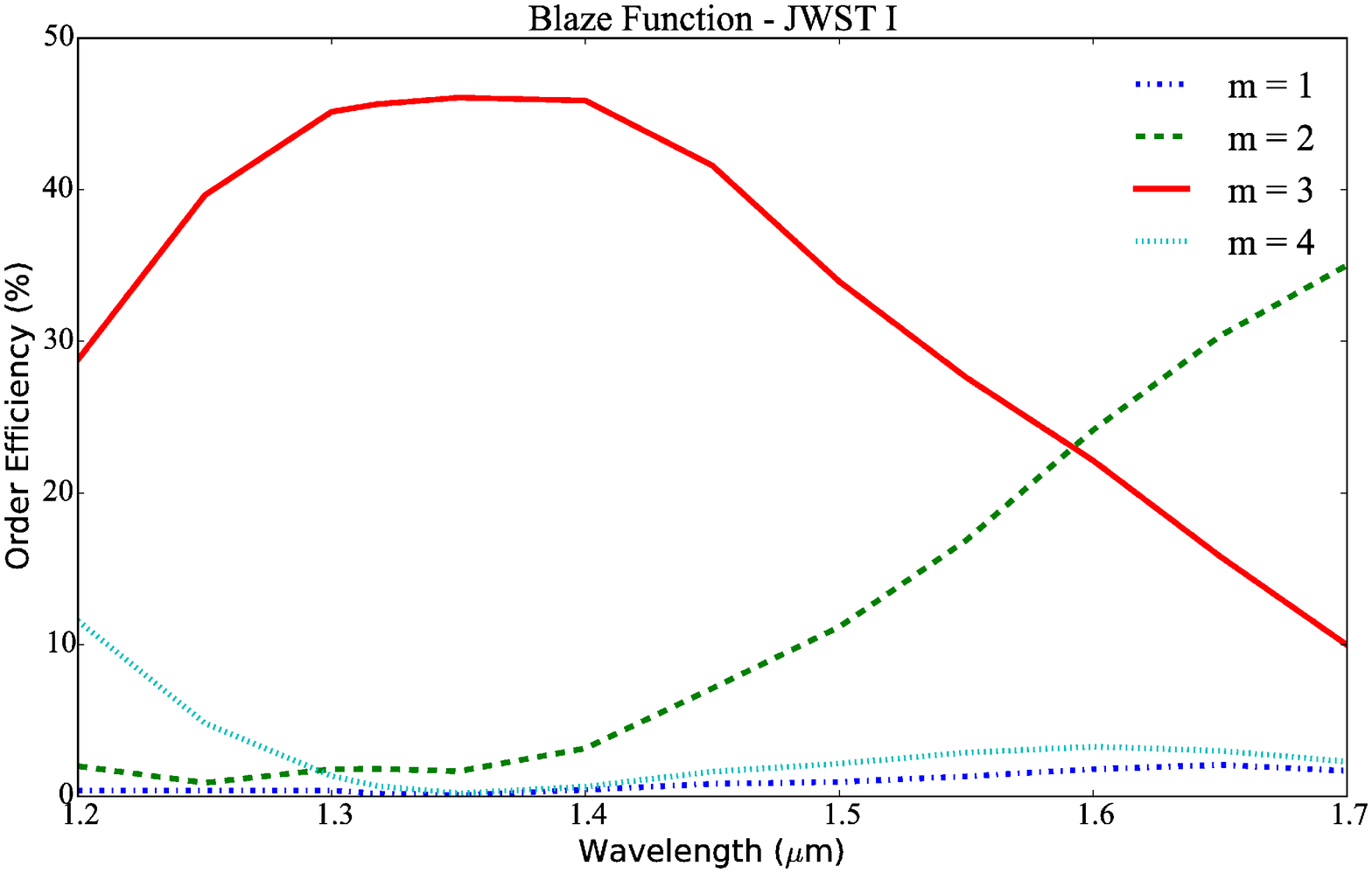}
  \end{center}
  \caption{Efficiency of grism A6I measured in third order.  Prior to coating, our test bench was not capable of measuring the blaze function in first order ($3.74 \mu$m), so instead we measured the blaze functions of higher orders in the regions where our test bench could function.}
   \label{fig:A6I_efficiency}
\end{figure}

In addition to the information about the trial part's efficiency in second order, we were
able to take efficiency measurements of the JWST flight part A6I before anti-reflection coating.
The grism-efficiency test bench available at the time only had sensitivity out to 1.8 $\mu$m.
We therefore measured the uncoated diffraction efficiency of A6I in third order (Figure 
\ref{fig:A6I_efficiency}).  The peak efficiency is 45\% at 1.35 $\mu$m for this uncoated part, in excellent agreement with the maximum possible transmission ($\sim 46\%$) for an uncoated silicon grism with an opening angle of $\delta=11.0^\circ$.

\section{Conclusions}
\subsection{Grism System Design}

The new generation of wide-field infrared imagers with large detector arrays offers a new
set of possibilities for imager-spectrometer conversions using grisms.  In particular, space-based
imagers can be very powerful as slitless transmission spectrographs.  On the ground, perhaps
the most significant change that the new instruments bring, when combined with grisms made 
from high index materials,  is a displacement of the resolving
power boundary where designers need to go over to dedicated spectrographs.  Previously, this
boundary lay at R$\sim$1000.  With cross-dispersed transmissive designs and Si grisms, 
designs with decent-sized slits, broad spectral grasp, and R approaching 10$^4$ become 
possible.  

\subsection{Ultimate Performance of Silicon Grisms}

Silicon grisms offer exceptional promise as the central element of transmission spectrographs 
in the infrared.  In the discussion above, we have shown that the losses due to groove placement
errors, entrance surface figure, and small-scale roughness are essentially negligible.  Measurements
of completed devices place some bounds on how well we can do.  The measured peak
efficiency of our double-side coated JWST trial part of 75\% provides one data point while the
peak efficiency of the uncoated flight part A6I of 45\% provides another.  It is worthwhile asking
what efficiency we could now reach, given the benefit of the experience with these parts.

One of the largest sources of loss for both parts is the blockage caused by the flat strips that
serve as etch-stops.  In the geometry of the JWST parts, these flats are responsible for an 
18\% loss.  Conventional UV contact lithography, however, cannot reliably produce strips
much narrower than the 1.4 $\mu$m ones used on these gratings.  If we turn, however,
to direct writing with an electron beam machine \citep{GullySantiago2014}, we can make usable etch stops as narrow as 200 nm.  At this width, the geometric
limitation on groove blockage in gratings with a 6$^o$ blaze comes from the opposite
side of the 70$^o$ vertices, rather than from the flat strips.  Based on Fig. 6 of 
Mar et al. (2009), the loss can be reduced to $\sim$7\%.  Given the ideal maximum 
efficiency of the trial part of 82\%, an antireflection coated grism with small flat strips
could have a peak efficiency of $\sim$85\% with existing coatings.  Further improvement
of the coatings, for example using fewer layers on the grooved side and trading
band width for uniformity, could result in even higher peak efficiencies.

The authors would like to thank our referee Stephen Shectman for insightful comments that improved the manuscript in terms of readability and accuracy.

\bibliography{bibliog}
\bibliographystyle{apj}

\end{document}